
\magnification=1200

\headline{\ifnum\pageno=1 \nopagenumbers
\else \hss\number \pageno \fi \hss}
\footline={\hfil}
\baselineskip=14pt
\overfullrule=0pt
\font\boldgreek=cmmib10
\textfont9=\boldgreek
\mathchardef\myalpha="090B
\def\bfalpha{{\fam=9 \myalpha}\fam=1}
\def\tvi{\vrule height 12pt depth 6pt width 0pt}
\def\tv{\tvi\vrule}
\def \cc #1 {\kern .7em \hfill #1 \hfill \kern .7em}

\centerline{\bf UNCERTAINTIES ON THE CP PHASE $\bfalpha$}\par
\centerline{\bf DUE TO PENGUIN DIAGRAMS}\par
\vskip 8 truemm
\centerline{R. Aleksan}
\centerline{DPhPE, CEN Saclay, 91191 Gif-sur-Yvette Cedex, France} \par
\vskip 5 mm
\centerline{F. Buccella}
\centerline{Dipartimento di Science Fisiche, Universit\`a degli Studi di
Napoli}
\centerline{Mostra d'Oltremare, Padiglione 19-20, 80125 Napoli, Italy}
\vskip 5 mm
\centerline{A. Le Yaouanc, L. Oliver, O. P\`ene and J.-C. Raynal} \par
\centerline{Laboratoire de Physique Th\'eorique et Hautes
Energies\footnote{*}{Laboratoire
associ\'e au Centre National de la Recherche Scientifique (URA 63)}} \par
\centerline{Universit\'e de Paris XI, b\^atiment 211, 91405 Orsay Cedex,
France} \par

\vskip 1 truecm
\noindent \underbar{{\bf Abstract}} \par
A major problem in the determination of the the CP angle $\alpha$, that should
be measured through
modes of the type $B_d, \bar{B}_d \to \pi \pi , \cdots ,$ is the uncertainty
coming from Penguin
diagrams. We consider the different ground state modes $\pi \pi$, $\pi \rho$,
$\rho \rho$, and,
assuming the FSI phases to be negligible, we investigate the amount of
uncertainty coming from
Penguins that can be parametrized by a dilution factor $D$ and an angle shift
$\Delta \alpha$. The
parameter $D$ is either 1 or very close to 1 in all these modes, and it can be
measured
independently, up to a sign ambiguity, by the $t$ dependence. Assuming
factorization, we show that
$\Delta \alpha$ is much smaller for the modes $\rho \pi$ and $\rho \rho$ than
for $\pi \pi$, and we
plot their allowed region as a function of $\alpha$ itself. Moreover, we show
that most of the modes
contribute to the asymmetry with the same sign, and define for their sum an
effective $D_{eff}$ and
an effective $\Delta \alpha_{eff}$, an average of $\Delta \alpha$ for the
different modes. It turns
out that $D_{eff}$ is of the order of 0.9, $\Delta \alpha_{eff}$ is between 5
$\%$ and 10 $\%$ and,
relative to $\pi \pi$, the statistical gain for the sum is of about a factor
10. Finally, we compute
the ratios $K\pi /\pi \pi , \cdots$ that test the strength of the Penguins and
depend on the CP
angles, as emphasized by Silva and Wolfenstein and by Deandrea et al.

 \par
\vskip 1 truecm
\noindent LPTHE Orsay 95-22 \par
\noindent hep-ph/9506260 \par
\noindent March 1995

\vfill \supereject
\baselineskip=20pt
We will adopt Wolfenstein phase convention$^{(1)}$ and parametrization of the
CKM matrix, with the
expansion in powers of $\lambda$ (up to order $\lambda^3$ included). In this
convention all CKM matrix
elements are real except $V_{ub}$ and $V_{td}$ and it is simple to identify
which modes will
contribute to the determination of the different angles on the unitarity
triangle $\alpha$, $\beta$
and $\gamma$. In the Standard Model we have $|q/p| = 1$ to a very good
approximation. In Wolfenstein
phase convention $(q/p)_{B_d}$ is complex since it depends on $V_{td}$  while
$(q/p)_{B_s}$ is
real as it depends on $V_{ts}$. For $B$ decays, the CKM factor of the decay
amplitudes is real for $b
\to c$ transitions while it is complex for $b \to u$ transitions. This gives us
four different
possibilities according to the value of $Im \left [ {q \over p} {\bar{M} \over
M} \right ]$~:
\par

1) $b \to u$ transitions of the $B_d$-$\bar{B}_d$ system, related to the angle
$\alpha$~; \par

2) $b \to c$ transitions of the $B_d$-$\bar{B}_d$ system, related to $\beta$~;
\par

3) $b \to u$ transitions of the $B_s$-$\bar{B}_s$ system, related to $\gamma$~;
\par

4)  $b \to c$ transitions of the $B_s$-$\bar{B}_s$ system, related to the angle
called $\beta '$,
vanishing at the considered order for the CKM matrix. Actually $\beta '$ is of
order $\lambda^2$.
\par

	Examples of the four types of modes, which are CP eigenstates, are
respectively~:  $B_d , \bar{B}_d
\to \pi^+ \pi^-$~; $B_d , \bar{B}_d \to \psi K_s, D^+D^-$~; $B_s , \bar{B}_s
\to \rho^0 K_s$, and
$B_s , \bar{B}_s \to \psi \varphi$. Of course, this is only true in the tree
approximation~:
{\it Penguin diagrams can complicate the picture and make uncertain the
determination of the angles},
in some channels. \par

A systematic study of the contribution of Penguin diagrams to CP asymmetries
has been done recently
by A. Deandrea et al.$^{(2)}$, together with the study of the decay rates of
modes where the Penguin
diagrams can be dominating, as in CKM suppressed modes like $\bar{B}^0 \to K^-
\pi^+$. Earlier
literature on the importance of Penguin diagrams in $B$ decays include the
works by Gavela et al.,
Guberina and Peccei, Eilam$^{(3)}$, and recently, Deshpande and Trampetic and
F. Buccella
and collaborators$^{(4)}$. On the other hand, Silva and Wolfenstein$^{(5)}$ and
Deandrea et
al.$^{(2)}$ have  recently emphasized that the ratios of the type $K \pi /\pi
\pi, \cdots$ give an
independent determination of the CP angles, rather free of hadronic
uncertainties if one assumes
factorization. However, one should keep in mind that this type of determination
does not involve CP
violation, and cannot make the economy of measuring CP asymmetries. \par

	The aim of the present paper is to examine which are the most advantageous
modes to determine the
angle $\alpha$ of the unitarity triangle as far as the Penguin diagram
uncertainty is concerned. This
is the angle where the Penguin contribution can be the most important source of
error. For the
case of the angle $\beta$, although the modes of the type $B_d$, $\bar{B}_d \to
D^+D^-, \cdots$ have
also Penguin contributions, one has the very clean mode $B_d$, $\bar{B}_d \to
\psi K_s$ which is, in
practice, clean of such contamination. As far as the determination of the angle
$\gamma$ is
concerned, one could in principle use the mode $B_s$, $\bar{B}_s \to \rho^0
K_s$. However, this is
unlikely to be feasible since this mode is not only CKM suppressed (order
$\lambda^3$ in amplitude),
but also color suppressed, which means still another suppression factor of the
order 0.2.
Thus, in the case of the angle $\gamma$ one should turn to modes of the type
$B_s$, $\bar{B}_s \to
K^-D^+_s, \cdots$ which are not CP eigenstates, and which are not affected by
Penguin diagrams. The
main hadronic uncertainties here concern the determination of the dilution
factor $D^{(6,7)}$. \par

	It has been pointed out by Gronau and London$^{(8)}$ that it could be possible
to separate the
Penguin contribution (pure $\Delta I=1/2$) from the tree contribution (that has
both $\Delta I = 1/2$
and $\Delta I = 3/2$ pieces) by isospin analysis of the different $\pi \pi$
channels. However,
although this is in principle possible up to discrete ambiguities, it seems
very difficult in
practice, essentially because not only class I$^{(9)}$ decays like $\pi^+
\pi^-$ are CKM suppressed,
but class II decays like $\pi^0 \pi^0$, color suppressed, have smaller
branching ratios by about two
orders of magnitude. Moreover, Deshpande and He$^{(10)}$ have recently pointed
out that the
Electroweak Penguins are not completely negligible in $B$ decays (one operator
has a sizeable Wilson
coefficient), invalidating the isospin analysis since this contribution does
not respect the usual
isospin properties. In particular, not surprisingly, this new contribution
enhances the relative
weight of Penguins in the supressed modes like $\pi^0 \pi^0$. \par

	Therefore, it seems sensible to start with the dominant class I decays and
investigate
the uncertainty in the determination of the angle $\alpha$ coming from the
Penguin contributions.
However, even here, there is still a further uncertainty coming from strong
(FSI)
phases. We shall neglect these possible FSI phases in this paper~: on the one
hand we do not know how
to predict them, and furthermore, we can expect that, for a heavy system like
the $B$ with light
decay products, they will be very small, since the final states will have large
velocities. For sure,
this point deserves further investigation. \par

We will restrict then to color allowed class I modes, since these will have the
larger branching
ratios, and moreover factorization is presumably for them on a rather firm
ground. A point of
warning must be made however. Since we are dealing with decays to light quarks,
the heavy-to-light
meson form factors at large momentum transfer will be involved, which are the
worst known.  \par

To summarize, we will consider the modes (with the different polarization
states) :
$$B_d \ , \ \bar{B}_d \to \pi^+ \pi^- \ , \ \pi^+ \rho^- \ , \ \rho^+ \pi^- \ ,
\ \rho^+ \rho^- \ \ \
. \eqno(1)$$

	Let us consider the final states :
$$\eqalignno{
|f > = &\left . \left | \pi^-({\bf p}) \pi^+(- {\bf p}) \right > \right . 	\cr
&\left . \left | \pi^-({\bf p}) \rho^+(\lambda = 0,- {\bf p}) \right >	\right.
\cr
&\left. \left | \rho^-(\lambda = 0, {\bf p}) \pi^+(-{\bf p}) \right >	\right .
\cr
&\left . \left |\rho^-(\lambda = 0, {\bf p}) \rho^+(\lambda = 0,- {\bf p})
\right >	\right . \cr
&\left . \left | \rho^-(\lambda = \pm , {\bf p}) \rho^+(\lambda = \pm,- {\bf
p}) \right > \right .
&(2) \cr  }$$

\noindent and their CP conjugate modes~:
$$\eqalignno{
|\bar{f}> = &\left . \left | \pi^+(- {\bf p}) \pi^-({\bf p}) \right > \right .
\cr
& - \left . \left | \pi^+(- {\bf p}) \rho^-(\lambda = 0, {\bf p}) \right >
\right . \cr
&- \left . \left | \rho^+(\lambda = 0,- {\bf p}) \pi^-({\bf p}) \right >
\right.	\cr
& \left . \left | \rho^+(\lambda = 0,- {\bf p}) \pi^-(\lambda = 0, {\bf p})
\right >	\right . \cr
&\left . \left | \rho^+(\lambda = \pm,- {\bf p}) \rho^-(\lambda = \pm, {\bf p})
\right > \right .
\ \ \ . &(3) \cr }$$

\noindent The spin quantization axis is along the line of flight of the decay
products in the $B_d$
rest frame. \par

	The effective Hamiltonian, following Buras et al.$^{(11)}$ is given
by$^{(2,10)}$~:
$$H =  {G \over \sqrt{2}}  \left \{ V^*_{ud}V_{ub} \left ( c_1 O_1 +
c_2 O_2 \right ) - V^*_{td} V_{tb} \left ( c_3 O_3 + c_4 O_4 + c_5 O_5 + c_6
O_6 \right ) \right \} +
H^{Penguin}_{EW} \eqno(4)$$

\noindent where the operators and Wilson coefficients are given by (at $\mu =
m_b$)~:

$$\matrix{
&O_1 = \left [ \bar{u} \gamma_{\mu}\left (1-\gamma_5 \right ) b \right ] \left
[ \bar{d} \gamma_{\mu}
\left ( 1 - \gamma_5 \right ) u \right ] \hfill		 &\qquad c_1 = 1.1502 \hfill
\cr
& & \cr
&O_2 = \left [ \bar{u}_{\alpha} \gamma_{\mu}\left (1-\gamma_5 \right )
b_{\beta} \right ] \left [
\bar{d}_{\beta} \gamma_{\mu} \left ( 1 - \gamma_5 \right ) u_{\alpha} \right ]
\hfill		 &\qquad c_2 =
- 0.3125 \hfill \cr & & \cr
&O_3 = \left [ \bar{d} \gamma_{\mu}\left (1-\gamma_5 \right ) b \right ] \left
[ \bar{q} \gamma_{\mu}
\left ( 1 - \gamma_5 \right ) q \right ] \hfill		 &\qquad c_3 = 0.0174 \hfill
\cr
& & \cr
&O_4 = \left [ \bar{d}_{\alpha} \gamma_{\mu} \left ( 1 - \gamma_5 \right )
b_{\beta} \right ]
\left [ \bar{q}_{\beta} \gamma_{\mu} \left ( 1 - \gamma_5 \right ) q_{\alpha}
\right ] \hfill	 &\qquad
c_4 = -0.0373 \hfill \cr
& & \cr
&O_5 = \left [ \bar{d} \gamma_{\mu} \left ( 1 - \gamma_5 \right ) b \right ]
\left [ \bar{q}
\gamma_{\mu} \left ( 1 + \gamma_5 \right ) q \right ]	\hfill	 &\qquad c_5 =
0.0104 \hfill \cr
& & \cr
&O_6 = \left [ \bar{d}_{\alpha} \gamma_{\mu} \left ( 1 - \gamma_5 \right )
b_{\beta} \right ]
\left [ \bar{q}_{\beta} \gamma_{\mu} \left ( 1 + \gamma_5 \right ) q_{\alpha}
\right ] \hfill	 &\qquad
c_6 = - 0.0459 \hfill \ \ \ . \cr } \eqno(5)$$

\vskip 5 mm
The Electroweak Penguin effective Hamiltonian $H^{Penguin}_{EW}$ has been
recently
computed by Deshpande and He, following Buras et al.$^{(12)}$. Only one
operator has a sizeable
coefficient~:
$$H^{Penguin}_{EW}  =  - {G \over \sqrt{2}}  V^*_{td} \ V_{tb} \left ( c_7 O_7
+
c_8 O_8 + c_9 O_9 + c_{10} O_{10} \right ) \eqno(6)$$

$$\matrix{
&O_7 = {3 \over 2} \left [ \bar{d} \gamma_{\mu} \left ( 1 - \gamma_5 \right ) b
\right ]
\left [ e_q \ \bar{q} \gamma_{\mu} \left ( 1 + \gamma_5 \right ) q \right ]
\hfill &\qquad c_7 = -1.050
\times 10^{-5} \hfill\cr
& & \cr
&O_{8} = {3 \over 2} \left [ \bar{d}_{\alpha} \gamma_{\mu} \left ( 1 - \gamma_5
\right ) b_{\beta} \right ] \left [ e_q \ \bar{q}_{\beta} \gamma_{\mu} \left (
1 + \gamma_5 \right )
q_{\alpha} \right ]	\hfill &\qquad c_{8} = 3.839 \times 10^{-4}  \hfill\cr
& & \cr
&O_9 = {3 \over 2} \left [ \bar{d} \gamma_{\mu} \left ( 1 - \gamma_5 \right ) b
\right ]
\left [ e_q \ \bar{q} \gamma_{\mu} \left ( 1 - \gamma_5 \right ) q \right ]
\hfill &\qquad c_9 =
-0.0101 \hfill \cr
& & \cr
&O_{10} = {3 \over 2} \left [ \bar{d}_{\alpha} \gamma_{\mu} \left ( 1 -
\gamma_5 \right ) b_{\beta}
\right ] \left [ e_q \ \bar{q}_{\beta} \gamma_{\mu} \left ( 1 - \gamma_5 \right
) q_{\alpha} \right ]
\hfill &\qquad c_{10} = 1.959 \times 10^{-3} \hfill \ \ \ . \cr
}\eqno(7)$$
\vskip 3 mm

	We will first make the calculation with the strong Penguin and then see that
things are only
very slightly modified by the electroweak one. \par

	From the definitions
\vfill \supereject
$$\eqalignno{
&\left < P(p) \left | A_{\mu} \right | 0 \right > = -i f_P \ p_{\mu} \cr
&\left < V(p,\lambda )\left | V_{\mu} \right | 0 \right >  = m_V f_V
\varepsilon_{\mu}^* (\lambda ) \cr
&\left < P_i \left | V_{\mu} \right | P_j \right > = \left ( p_i^{\mu} +
p_j^{\mu} -{m^2_j - m^2_i
\over q^2} q_{\mu} \right ) f_+(q^2) + {m^2_j - m^2_i \over q^2}  q_{\mu}
f_0(q^2) \cr
&\left < V_i \left | A_{\mu} \right | P_j \right > = i \left ( m_i + m_j \right
)
A_1(q^2) \left ( \varepsilon_{\mu}^* - {\varepsilon^* \cdot q \over q^2}
q_{\mu} \right ) - \cr
&- i A_2(q^2) {\varepsilon^* \cdot q \over m_i + m_j} \left ( p_i^{\mu} +
p_j^{\mu} - {m^2_j - m^2_i
\over q^2}  q_{\mu} \right ) + 2 i m_i A_0(q^2)  {\varepsilon^* \cdot q \over
q^2}  q_{\mu} \cr
&\left < V_i \left | V_{\mu} \right | P_j \right > = {2 V(q^2) \over m_i + m_j}
  \varepsilon_{\mu
\nu \rho \sigma} \ p^{\nu}_j \ p^{\rho}_i \ \varepsilon^{*\sigma} &(8) \cr
}$$

\noindent we obtain, neglecting for the moment the Electroweak Penguin, the
expressions~:
$$\eqalignno{
&M \left ( \bar{B}^0_d \to \pi^- ({\bf p}) \pi^+ (-{\bf p}) \right ) =  {G
\over \sqrt{2}}  i
f_{\pi} \left ( m^2_B - m^2_{\pi} \right )  f_0^{ub}(m^2_{\pi}) \cr
&\times \left ( V^*_{ud} V_{ub} a_1 - V^*_{td} V_{tb} \left \{ a_4 + a_6
{2M^2_{\pi} \over \left ( m_b
- m_u \right ) \left ( m_u + m_d \right )} \right \} \right ) \cr
&  \cr
&M \left ( \bar{B}^0_d \to \rho^- (\lambda = 0, {\bf p}) \pi^+ (-{\bf p})
\right ) = {G \over
\sqrt{2}} \ 2 f_{\rho} \ m_B \ f_+^{ub}(m^2_{\rho}) p \cr
& \times \left ( V^*_{ud} V_{ub} \ a_1 - V^*_{td} V_{tb} \ a_4 \right ) \cr
& \cr
&M \left ( \bar{B}^0_d \to \pi^- ({\bf p}) \rho^+ (\lambda = 0,-{\bf p}) \right
) =  - {G \over
\sqrt{2}} \  2 f_{\pi} \ m_B \ A_0^{ub}(m^2_{\pi}) p \cr
&\times \left ( V^*_{ud} V_{ub} \ a_1 - V^*_{td} V_{tb} \left \{ a_4 - a_6
{2M^2_{\pi} \over \left (
m_b - m_u \right ) \left ( m_u + m_d \right )} \right \} \right ) \cr
& \cr
&M \left ( \bar{B}^0_d \to  \rho^- (\lambda = 0, {\bf p}) \rho^+ (\lambda =
0,-{\bf p}) \right ) = i
{G \over \sqrt{2}} \  m_{\rho} \ f_{\rho} \cr
&\left [ \left ( m_B + m_{\rho} \right ) \left ( {2p^2 + m^2_{\rho} \over
m^2_{\rho}} \right )
A_1^{ub}(m^2_{\rho}) - {m^2_B \over m_B + m_{\rho}}  {2p^2  \over m^2_{\rho}}
A_2^{ub}(m^2_{\rho})
\right ] \cr
&\times \left ( V^*_{ud} V_{ub} \ a_1 - V^*_{td} V_{tb} \ a_4 \right ) \cr
& \cr
&M^{pv} \left ( \bar{B}^0_d \to \rho^- (\lambda =\pm,{\bf p}) \rho^+ (\lambda =
\pm,-{\bf p}) \right )
= i {G \over \sqrt{2}}  m_{\rho} f_{\rho} \left ( m_B + m_{\rho} \right )
A_1^{ub}(m^2_{\rho}) \cr
& \times \left ( V^*_{ud} V_{ub} \ a_1 - V^*_{td} V_{tb} \ a_4 \right ) \cr
& \cr
&M^{pc} \left ( \bar{B}^0_d \to \rho^- (\lambda = \pm, {\bf p}) \rho^+ (\lambda
= \pm,-{\bf p}) \right
) = \pm  i {G \over \sqrt{2}}  m_{\rho} f_{\rho}  {m_B \over m_B + m_{\rho}} 2
V^{ub}(m^2_{\rho}) p
\cr &\times \left ( V^*_{ud} V_{ub} \ a_1 - V^*_{td} V_{tb} \ a_4 \right ) &(9)
\cr
}$$

\noindent where we have used the notation of Deandrea et al.$^{(2)}$
$$a_{2i-1} = c_{2i-1} + {c_{2i} \over N_c} 		\qquad a_{2i} = c_{2i} + {c_{2i-1}
\over N_c}
\qquad \qquad (i = 1,2,3) \ \ \ . \eqno(10)$$

As far as the combination of Wilson coefficients is concerned, these
expressions agree with the ones
given in the Tables of the paper by Deandrea et al.$^{(2)}$. Let us make a
comment on the
Electroweak Penguins (6). For the modes under consideration the matrix elements
of the operators
$O_{7}$, $O_{8}$, $O_{9}$ and $O_{10}$ are equal to those of respectively the
operators $O_{5}$,
$O_{6}$, $O_{3}$ and $O_{4}$. Therefore, the inclusion of the Electroweak
Penguins simply makes the
changes $a_4 \to a_4 + a_{10}$ and $a_6 \to a_6 + a_8$ in the preceding
formulae, that amounts to
changes only at the percent level in these coefficients. Notice that Deshpande
and He find
significant changes due to the Electroweak Penguins in the case of the $\pi^0
\pi^0$, ... modes that
are color suppressed and that we do not consider here.  \par

Let us write the relevant asymmetries in our case for a definite type of final
state like $|f> =
|\rho^{-}(\lambda = 0,{\bf p})\pi^{+}(-{\bf p})>$ and its CP conjugate mode
$|\bar{f}> =
-|\pi^{-}({\bf p})\rho^+(\lambda = 0,-{\bf p})>$. \par

	The amplitudes will write, splitting into Tree and Penguin Amplitudes
(remember that we neglect FSI
phases)~: $$\eqalignno{
&\bar{M}(f) = M_T \ e^{i\beta_T} + M_P \ e^{i\beta_P} \equiv M_1 \ e^{i\beta_1}
\cr
&\bar{M}(\bar{f}) = \bar{M}_T \ e^{i\beta_T} + \bar{M}_P \ e^{i\beta_P} \equiv
M_2 \ e^{i\beta_2} \cr
&M(\bar{f}) = - \left [ M_T \ e^{-i\beta_T} + M_P \ e^{-i\beta_P} \right ]
\equiv - M_1
\ e^{-i\beta_1} \cr
&M(f) = - \left [ \bar{M}_T \ e^{-i\beta_T} + \bar{M}_P \ e^{-i\beta_P} \right
] \equiv - M_2
\ e^{-i\beta_2} &(11) \cr
}$$

\noindent ($M_1$ and $M_2$ are the moduli of $\bar{M}(f)$ and of
$\bar{M}(\bar{f})$).
 \par

	Let us define $\Delta \alpha$ and $n_f$ through~:
$${q \over p} \ {\bar{M}(f) \over M(f)}  \cong - {M_1 \over M_2} \
e^{i(\varphi_M + \beta_1 +
\beta_2)} \equiv - \eta_f {M_1 \over M_2} \  e^{2i(\alpha + \Delta \alpha)}
\eqno(12)$$

\noindent where $\alpha$ is the angle of the unitarity triangle, and $\eta_f$
is a sign that depends
on the final state. As we will see below, $\eta_f = +1$ for $\pi^+ \pi^-$,
$\rho^+ \rho^-$ (parity
violating part), $\pi^+ \rho^- + \pi^- \rho^+$, and $\eta_f = -1$ for $\rho^+
\rho^-$ (parity
conserving part). The correction $\Delta \alpha$ ($\Delta \alpha \to 0$ for
$M_P \to 0$) is given
by~:  $$\Delta \alpha = {1 \over 2} \left ( \beta_1 + \beta_2 \right ) -
\beta_T \ \ \ . \eqno(13)$$

	The time-dependent rates are proportional to~:
$$\vbox{\eqalignno{
&R \left ( B_{phys}^0(t) \to f \right ) \sim \left [ 1 - R \cos (\Delta Mt) - D
\sin [2(\alpha +
\Delta \alpha )] \sin(\Delta Mt) \right ] \cr
&R \left ( \bar{B}_{phys}^0(t) \to f \right ) \sim \left [1 + R \cos (\Delta
Mt) + D \sin
[2(\alpha + \Delta \alpha )] \sin( \Delta Mt) \right ] \cr
&R \left ( B_{phys}^0(t) \to \bar{f} \right ) \sim \left [ 1 + R \cos (\Delta
Mt) - D \sin
[2(\alpha + \Delta \alpha	)] \sin (\Delta Mt) \right ] \cr
&R \left ( \bar{B}_{phys}^0(t) \to \bar{f} \right ) \sim \left [ 1 - R \cos
(\Delta Mt) + D
\sin [2(\alpha + \Delta \alpha )] \sin (\Delta Mt) \right ] &(14) \cr }}$$

\noindent where
$$R = {(M_1)^2 - (M_2)^2 \over (M_1)^2 + (M_2)^2}			\qquad D = {2M_1M_2 \over
(M_1)^2 + (M_2)^2}
\ \ \ . \eqno(15)$$

\noindent From these expressions, a useful CP asymmetry that we can consider~:
$$\left \{ R \left ( B_{phys}^0(t) \to f \right ) + R \left ( B_{phys}^0(t) \to
\bar{f} \right ) \right \} -  \left \{ R \left ( \bar{B}_{phys}^0(t) \to f
\right ) +
R \left ( \bar{B}_{phys}^0(t) \to \bar{f} \right ) \right \}  \sim$$  $$\sim D
\sin [2(\alpha + \Delta
\alpha )] \sin (\Delta Mt) \ \ \ . \eqno(16)$$

\noindent We can ask two interesting questions~: \par \vskip 3 mm

1) By how much does $\alpha + \Delta \alpha$ differ from $\alpha$, the angle of
the unitarity
triangle~? $\Delta \alpha$ is a function of ${V^*_{td} \over V_{ub}} = {1 -
\rho +i \eta \over \rho -
i \eta}$.
 \par \vskip 3 mm

2) By how much does $D$ differ from 1~? For the non-CP eigenstates $D$ given by
(15) depends also on
$(\rho ,\eta)$. There are two kinds of uncertainties on $D$, then : i) the one
coming from the poor
knowledge of the form factors involved in the calculation, and ii) the one
coming from the Penguin
contribution. However, at least in principle, one can have independent
experimental information on $D$
(up to a sign ambiguity since $|D| = \sqrt{1 - R^2}$) from the study of the
time dependence
through the $\cos(\Delta Mt)$ in the formulae above. \par

Therefore, this second question is not as crucial as the first one, and the
important uncertainty
concerns $\Delta \alpha (\rho ,\eta)$. \par

As we have shown in refs. 13 and 7, it is possible to consider the CP asymmetry
for the whole sum
$\pi^+ \pi^- + \pi^+ \rho^- + \rho^+ \pi^- + \rho^ + \rho^-$. Let us consider
the sign of the
asymmetry for the different contributions, relative to the one for $\pi^+
\pi^-$ (parity violating,
$S$ wave, CP = +). In the case $\rho^+ \rho^-$ (see the formulae (9)), we have
the parity-violating
piece contributing to both the longitudinal and transverse amplitudes ($S$ and
$D$ waves, $CP = +$),
with therefore the same sign as $\pi^+ \pi^-$, while the parity-violating piece
($P$ wave, $CP = -$)
contributes with opposite sign. For the case $\pi^+ \rho^- + \rho^+ \pi^-$
(parity conserving, $P$
wave), the situation is more involved. Let us consider only the spectator
diagram, as we have done
writting formulae (9). We can neglect the exchange diagram because it  is color
suppressed, vanishes
in the chiral limit (current conservation if one assumes factorization), and
moreover has further form
factor suppressions, as discussed in ref. 13. We have shown in ref. 13 that in
the heavy quark limit
($c$ quark, as well as $b$ quark), the process $B_d, \bar{B}_d \to (D^+D^*- +
D^-D^{*+})(CP = +)$ is
only allowed by the spectator diagram, while $B_d, \bar{B}_d \to (D^+D^{*-} +
D^-D^{*+})(CP = -)$ is
only allowed by the exchange diagram, which is very small. As emphasized in
ref. 7, lattice
calculations do not show changes in sign for the form factors and decay
constants when extrapolating
from heavy to light masses. Assuming that this is the case for all form factors
involved, the
processes $B_d, \bar{B}_d \to (\pi^+ \rho^- + \rho^+ \pi^-)(CP = +)$ and $B_d,
\bar{B}_d \to
(\pi^+ \rho^- + \rho^+ \pi^-)(CP = -)$ will be allowed by the spectator
diagram, but with a
cancellation between both contributions ($\pi$ emission or $\rho$ emission) in
the latter, and a
constructive interference in the former. Hence, $B_d, \bar{B}_d \to \pi^+
\rho^- + \rho^+ \pi^-$ will
contribute to the asymmetry with the same sign as $\pi^+ \pi^-$ although with a
dilution factor coming
from the difference

$${\left [ \left ( A_{\pi} + A_{\rho} \right )^2 - \left ( A_{\pi} - A_{\rho}
\right )^2
\right ] \over \left [ \left ( A_{\pi} + A_{\rho} \right )^2 + \left ( A_{\pi}
- A_{\rho}
\right )^2 \right ]} = {2A_{\pi} A_{\rho} \over (A_{\pi})^2 + (A_{\rho})^2}
\eqno(17)$$

\noindent where $A_{\pi}$ and $A_{\rho}$ are the moduli of the amplitudes for
$\pi$ and $\rho$
emission respectively.  \par

For the whole sum$^{(1)}$, we will have, making a linear approximation on the
corrections (see ref.
13)~:  $$A(t) = D_{eff} \sin 2 \left [ \alpha + (\Delta \alpha)_{eff} \right ]
\sin \Delta Mt
\eqno(18)$$

\noindent with
$$D_{eff} = {2 \displaystyle{\sum_i} \eta_i M_1^{(i)} M_2^{(i)} p_i \over
\displaystyle{\sum_i}
\left [ \left ( M_1^{(i)} \right )^2 + \left ( M_2^{(i)} \right )^2 \right ]
p_i}		 \qquad \Delta
\alpha _{eff} \cong \sum_i D_i B_i(\Delta \alpha)_i \eqno(19)$$

\noindent where the sum extends over all modes enumerated above with momenta
$p_i$ and branching
ratios $B_i$. Of course, for the CP eigenmodes, $M_1 = M_2$. These amplitudes
can be read from (9) and
(11). Notice that only the $P$-wave parity conserving $\rho \rho$ mode
contributes negatively to the
numerator in this expression. In the expression of $\Delta \alpha_{eff}$, $D_i$
are the individual
dilution factors and $R_i$ the corresponding branching ratios relative to the
total sum of ground
state modes, given in the Table 1. We obtain finally~:
$$\Delta \alpha _{eff} = Arg \left [ 1 + 0.029 {1 -
\rho + i \eta	\over \rho - i\eta} \right ] \eqno(20)$$

\noindent essentially because the $\rho \rho$ modes dominate. \par

	In Table I we summarize the results, giving the branching ratios for the
different modes, and the
dilution factors and corresponding $\Delta \alpha$. We use the value
${2M^2_{\pi} \over (m_b - m_u) (m_u + m_d)} = 0.55$ for $m_u + m_d$ = 15 MeV
and $m_b =$ 4.7 GeV,
$${V^*_{td} V_{tb} \over V^*_{ud} V_{ub}} = {1 - \rho + i \eta \over \rho - i
\eta} =
-\left |{1 - \rho + i \eta \over \rho - i \eta} \right | e^{-i \alpha} =
- \left | {V_{td} \over V_{ub}} \right | e^{-i \alpha} \ \ \ . \eqno(21)$$

\noindent Notice that $\Delta \alpha$ for individual modes is independent of
the heavy-to-light
form factors. \par

In Fig. 1 we plot the allowed region in the plane ($\rho , \eta$) taking into
account present
theoretical and experimental uncertainties$^{(14)}$ and the corresponding
allowed domain in
the plane ($\sin 2 \alpha , \sin 2 \beta$). In Figs. 2 we plot $\Delta \alpha$
for the different modes
and for their sum as a function of $\sin 2 \alpha$ itself~: $\sin 2 \alpha$ and
$\Delta
\alpha$ as well are both functions of ($\rho , \eta$). \par

	The rates have been computed by extracting the $B \to \pi (\rho )$ form
factors from
the data on the $D$ semileptonic form factors $D \to K(K^*)$ at $q^2 =
0^{(15)}$~:
$$\eqalignno{
&f^{sc}_+(0) = 0.77 \pm 0.04 \cr
&V^{sc}(0) = 1.16 \pm 0.16 \cr
&A^{sc}_1(0) = 0.61 \pm 0.05 \cr
&A^{sc}_2(0) = 0.45 \pm 0.09 \ \ \ . &(22) \cr
}$$

	We use the following prescriptions, motivated by our study of the data on an
overall fit to these
$D$ semileptonic data and on $B \to \psi K(K^*)^{(16)}$ and our model of
semileptonic heavy meson form
factors$^{(17)}$~:  \par \vskip 3 mm

1) Let us begin by using exact SU(3) to relate $D \to K(K^*)$ to $D \to \pi
(\rho )$ at $q^2 = 0$.
Below we will see the sensitiveness to this hypothesis. It must be pointed out
that SU(3) is just
a simplifying assumption to get information on the $D \to \pi (\rho )$ form
factors. Preliminary
data on the ratio $D \to \pi/D \to K$ is consistent with SU(3) but also large
SU(3) breaking is
allowed by the present error. \par \vskip 3 mm

2) We extrapolate the form factors $D \to \pi (\rho )$
from $q^2 = 0$ to $q^2 = q^2_{max}$ by a $q^2$ extrapolation, taking
$$\vbox{\eqalignno{ &{f_0(q^2)
\over f_0(0)} = {A_1(q^2) \over A_1(0)} = 1 \cr &{f_+(q^2) \over f_+(0)} = {1
\over \left [ 1 - {q^2
\over \left ( M_D + M_{\pi} \right )^2} \right ]} \cr
&{A_0(q^2) \over A_0(0)} = {A_2(q^2) \over A_2(0)} = {V(q^2) \over V(0)} = {1
\over \left [ 1 -
{q^2 \over \left ( M_D + M_{\rho} \right )^2} \right ]} \ \ \ . &(23) \cr
}}$$

\noindent For the unmeasured form factors we take $f_0(0) = f_+(0)$ and $A_0(0)
= A_1(0)$. \par
\vskip 3 mm

3) We extrapolate the $D \to \pi (\rho )$ form factors at their $q^2_{max}$ to
the $B \to
\pi (\rho )$ form factors at their $q^2_{max}$ by using the softened
heavy-to-light
scaling laws$^{(15)}$~:
$$\eqalignno{
&{f^{db}_+(q^2_{max}) \over f^{dc}_+(q^2_{max})} = \left ( {M_D \over M_B}
\right )^{{1 \over 2}}
\left ( {M_B + M_{\pi} \over M_D + M_{\pi}} \right ) \cr
&{f^{db}_0(q^2_{max}) \over f^{dc}_0(q^2_{max})} = \left ( {M_B \over M_D}
\right )^{{1 \over 2}}
\left ( {M_D + M_{\pi} \over M_B + M_{\pi}} \right ) \cr
&{A^{db}_0(q^2_{max}) \over A^{dc}_0(q^2_{max})} = {A^{db}_2(q^2_{max}) \over
A^{dc}_2(q^2_{max})}
= {V^{db}(q^2_{max}) \over V^{dc}(q^2_{max})} = \left ( {M_D \over M_B} \right
)^{{1 \over 2}} \left (
{M_B + M_{\rho} \over M_D + M_{\rho}} \right ) \cr
&{A^{db}_1(q^2_{max}) \over A^{dc}_1(q^2_{max})} = \left ( {M_B \over M_D}
\right )^{{1 \over
2}} \left ( {M_D + M_{\rho} \over M_B + M_{\rho}} \right ) &(24) \cr
}$$

4) Finally we extrapolate the $B \to \pi (\rho)$ form factors from $q^2_{max}$
to $q^2 = M^2_{\pi}$ or $M^2_{\rho}$ by using pole relations (23), except for
the obvious
replacement $m_D \to m_B$. \par \vskip 3 mm

To test the stability of the results on our assumptions, we have made a number
of changes. Starting
from the central values (22) and extrapolating following the prescription
described above$^{(16)}$,
we have used exact SU(3) alternatively at $q^2= q^2_{max}$ for $D \to K(K^*)$
or at $q^2 = 0$
for $B \to K(K^*)$ or at $q^2= q^2_{max}$ for $B \to K(K^*)$. The results are
rather stable.
For example, the rate $B_d \to \pi^+ \pi^-$ changes by 20 $\%$, and $D_{eff}$
for the whole sum by 6
$\%$ (it increases). If instead of the central values (22) we adopt the best
fit to $D \to K(K^*)$
semileptonic form factors and $B \to \psi K(K^*)$ decay rates$^{(16)}$, we
obtain results that are
very close to the previous ones. \par

Finally, as a consistency test and to have an independent estimation of the
magnitude of the Penguin
diagrams, we will consider CP conserving processes, CKM suppressed modes where
the Penguin can be
dominant and which have the same topology as the modes interesting for CP
violation discussed above.
Their relative magnitude will be a precise test of the magnitude of the
Penguins. The formulae for the
amplitudes $M(\bar{B}^0_d \to K^- \pi^+)$, $M(\bar{B}^0_d \to K^{*-} \pi^+)$,
$M(\bar{B}^0_d \to K^-
\rho^+)$, $M(\bar{B}^0_d \to K^{*-} \rho^+)$, $M^{pv}(\bar{B}^0_d \to
K^{*-}(\lambda = \pm )
\rho^+(\lambda = \pm))$, $M^{pc}(\bar{B}^0_d \to K^{*-}(\lambda = \pm)
\rho^+(\lambda = \pm))$ can be
obtained from (9) by simply replacing $V_{ud}$ and $V_{td}$ by $V_{us}$ and
$V_{ts}$, and, when they
are emitted, $\pi$ and $\rho$ by $K$ and $K^*$. \par

We find, taking for all form factors $F^{ub}(m^2_K) = F^{ub}(m^2_{\pi})$,
$F^{ub}(m^2_{K^*}) =
F^{ub}(m^2_{\rho})$, but keeping SU(3) breaking for $f_K/f_{\pi}$, etc., and
adopting the value
${2M^2_K \over (m_b - m_u)(m_u + m_s)}  = 0.67$ ($m_s =$ 150 MeV)~:
$$\eqalignno{
&{\Gamma \left ( \bar{B}^0_d \to \pi^ + K^- \right ) \over \Gamma \left (
\bar{B}^0_d \to \pi^+ \pi^-
\right )} = 0.07 \ {|\rho - i\eta - 1.12|^2 \over |\rho - i \eta + 0.05 (1 -
\rho - i \eta )|^2} \cr
&{\Gamma \left ( \bar{B}^0_d
\to \pi^+ K^{*-} \right ) \over \Gamma \left ( \bar{B}^0_d \to \pi^+ \rho^{-}
\right )} = 0.06 \
{|\rho - i \eta -0.59|^2 \over |\rho - i \eta + 0.03 (1 - \rho + i \eta )|^2}
\cr
&{\Gamma \left ( \bar{B}^0_d \to \rho^+ K^-
\right ) \over \Gamma \left ( \bar{B}^0_d \to \rho^+ \pi^- \right )} = 0.07 \
{|\rho -i \eta -
0.05|^2 \over |\rho - i \eta + 0.01 (1 - \rho + i\eta )|^2} \cr
&{\Gamma \left ( \bar{B}^0_d \to \rho^+ K^{*-} \right ) \over \Gamma
\left ( \bar{B}^0_d \to \rho^+ \rho^- \right )} = 0.06 \ {|\rho - i \eta -
0.59|^2 \over |\rho - i
\eta + 0.03 (1 - \rho + i \eta )|^2} \ \ \ . &(25) \cr  }$$

In Fig. 3 we plot these ratios as a function of $\sin 2 \alpha$ and in Fig.~4
as a function of $\sin
2 \beta$. We see that, even considering the present uncertainties on the
coordinates ($\rho , \eta$),
these rates are rather sensitive to the precise value of $\sin 2 \beta$. Low
values of these ratios
could also give precious information on $\sin 2 \alpha$. Needless to say, the
measurement of these
ratios would be a consistency test (modulo the factorization approximation) but
cannot make the
economy of measuring directly the CP angles through CP violating processes.
\par

In conclusion, we have shown that, neglecting FSI phases, the uncertainties
$\Delta \alpha$ coming
from Penguin diagrams are smaller for the modes $\rho \pi$ and $\rho \rho$
than for $\pi \pi$. Moreover, summing over all these modes leads to an
uncertainty
$\Delta\alpha/\alpha$ of the order 5 to 10 $\%$. The dilution factor is very
close to 1 even for the
whole sum and one wins an order of magnitude in statistics. However, one must
keep in mind that these
results are sensitive to the Ansatz for the heavy-to-light meson form factors.
Our model relies on a
combination of theoretical constraints and data on semileptonic $D$ mesons
decays (subject to
corrections of the order $1/m_c$) and on non-leptonic $B$ decays (here the
serious uncertainty comes
from the factorization hypothesis). Needless to say that the dependence of the
form factors
on masses and $q^2$ could be quite different from our expectations, even if the
latter
stems from an extensive study$^{(16)}$. Therefore, the present analysis must be
considered as
preliminary. Further knowledge on the many uncertainties involved ($q^2$ and
mass dependence of
heavy-to-light form factors, FSI phases, accuracy of factorization) should be
included in future
analyses along the same lines.

\vskip 5 mm
\noindent {\bf Acknowledgements~:} This work has been supported in part by the
CEC Science
Project SCI-CT91-0729, and supported by the Human Capital and Mobility
Programme, contract
CHRX-CT93-0132.

\vfill \supereject
\centerline{\bf References}
\vskip 5 mm
\item{1.} L. Wolfenstein, Phys. Rev. Lett. {\bf 51}, 1945 (1983).
\item{2.} A. Deandrea, N. Di Bartolomeo, R. Gatto, F. Feruglio and N. Nardulli,
Phys. Lett. {\bf
B320}, 170 (1994)~; see also D. London and R.D. Peccei, Phys. Lett. {\bf B223},
257 (1989).
\item{3.} M.B. Gavela et al., B. Guberina, R. Peccei and R. R\"uckl, Phys.
Lett. {\bf B90}, 169
(1980)~; G. Eilam, Phys. Rev. Lett. {\bf 49}, 1478 (1982)~; M.B. Gavela et al.
Phys. Lett. {\bf
B154}, 425 (1985).
\item{4.} N.G. Deshpande and J. Trampetic, Phys. Rev. {\bf D41}, 895 (1990)~;
F. Buccella, G.
Mangano and G. Miele, Nuovo Cimento {\bf A104}, 1293 (1991)~; F. Buccella, G.
Mangano, G. Miele and
P. Santorelli, Nuovo Cimento {\bf A105}, 33 (1992).  \item{5.} J.P. Silva and
L. Wolfenstein, Carnegie
Mellon preprint, September 1993.   \item{6.} R. Aleksan, I. Dunietz and B.
Kayser, Zeit. Phys. {\bf
C54}, 653 (1992).  \item{7.} R. Aleksan, A. Le Yaouanc, L. Oliver, O. P\`ene
and J.-C. Raynal, Orsay
preprint LPTHE 94/03, to appear in Zeit. Phys. C.    \item{8.} M. Gronau and D.
London, Phys. Rev.
Lett. {\bf 65}, 3381 (1990).   \item{9.} M. Wirbel, B. Stech and M. Bauer,
Zeit. Phys. {\bf C29}, 637
(1985)~; {\bf C34}, 103 (1987).
\item{10.} N.G. Deshpande and Xiao-Gang He, University of Oregon
preprint OITS-553 (1994).
\item{11.} A.J. Buras, M. Jamin, M.E. Lautenbacher and P.H. Weisz, Nucl. Phys.
{\bf B370}, 69 (1992).
\item{12.} A.J. Buras, M. Jamin, M.E. Lautenbacher and P.H. Weisz, Nucl. Phys.
{\bf B400}, 37
(1993)~; A.J. Buras, M. Jamin and M.E. Lautenbacher, Nucl. Phys. {\bf B400}, 75
(1993).
\item{13.} R. Aleksan, A. Le Yaouanc, L. Oliver, O. P\`ene and J.-C. Raynal,
Phys. Lett. {\bf B317}, 173
(1993).
\vfill \supereject
\item{14.} We have taken, as it is usually done, $\left | {V_{ub} \over V_{cb}}
\right | = 0.08 \pm
0.02$. But we believe that this underestimates the theoretical error. Indeed,
this number is
extracted from a comparison with experiment~: i) with quark models that differ
from one another up
to a factor 2, and ii) with a parton model, for which it is impossible to bound
the
non-perturbative corrections near the end-point. In Figure 1, the bound from
$\varepsilon_K$ is
estimated from $0.6 < B_K < 1$ as indicated by lattice calculations. A lower
value of $B_K$ as
indicated by chiral Perturbation Theory is not excluded. (J. F. Donoghue, E.
Golowich and B. R.
Holstein, Phys. Lett. {\bf 119B}, 412 (1982)~; the corrections have been
calculated by C. Bruno,
Phys. Lett. {\bf B320}, 135 (1994)).    \item{15.} M.S. Witherell, Invited talk
given at the
International Symposium on Lepton and Photon Interactions at High Energies,
Cornell University,
Ithaca, N.Y. (1993), UCSB-HEP-93-06.    \item{16.} R. Aleksan, A. Le Yaouanc,
L. Oliver, O. P\`ene and
J.-C. Raynal, preprint DAPNIA/SPP 94-24 and LPTHE 94/15, hep-ph/9408215, to
appear in Phys. Rev. D.
\item{17.} A. Le Yaouanc, L. Oliver, O. P\`ene and J.-C. Raynal, to appear.

\vfill \supereject
\centerline{\bf Figure Captions}
\vskip 5 mm
{\parindent = 1.5 truecm
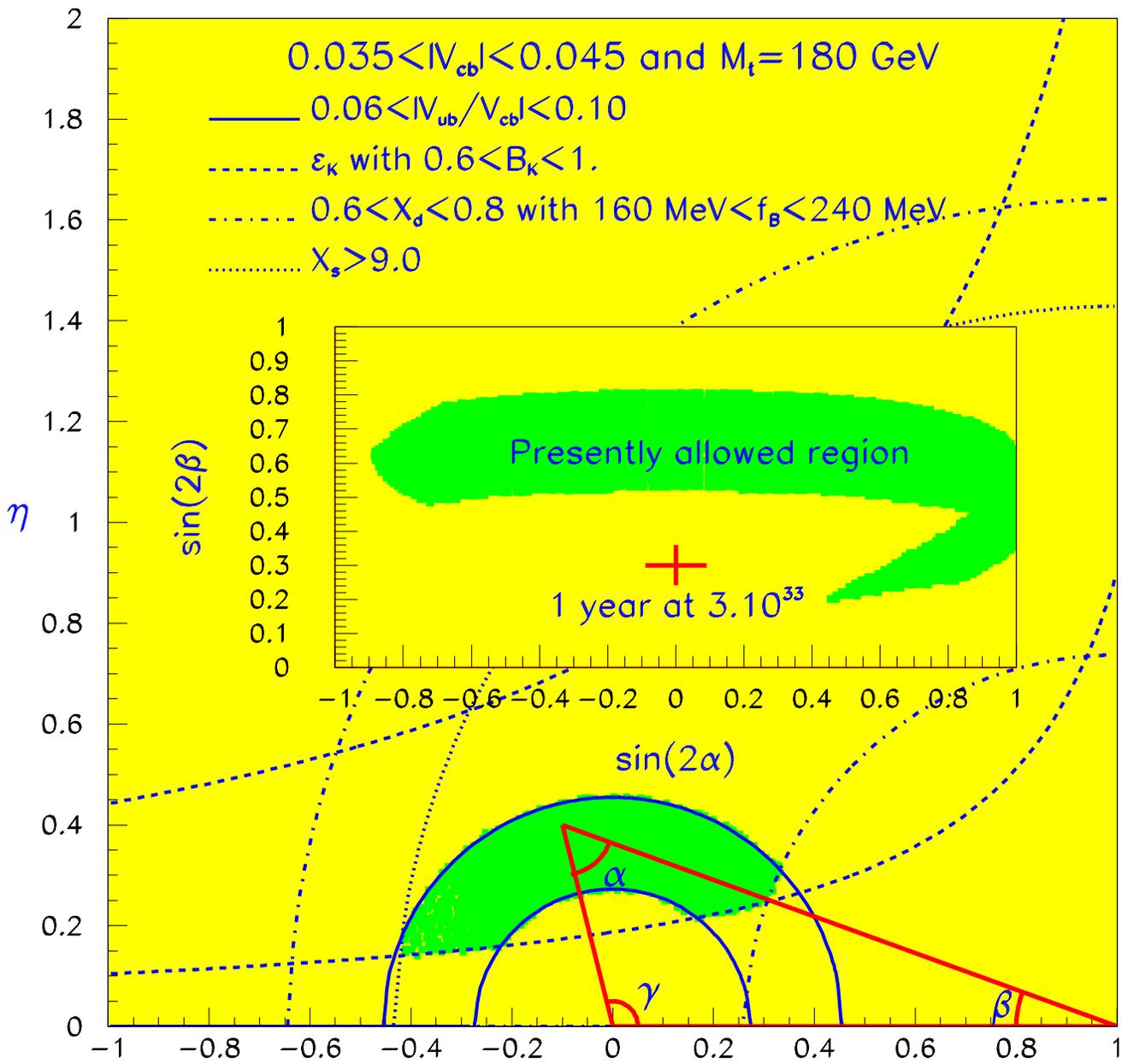
\item{\bf Fig. 1} Present allowed domain for the coordinates ($\rho, \eta$) and
for ($\sin 2 \alpha$, $\sin 2 \beta$). The theoretical
and experimental uncertainties are specified.
\vskip 3 mm

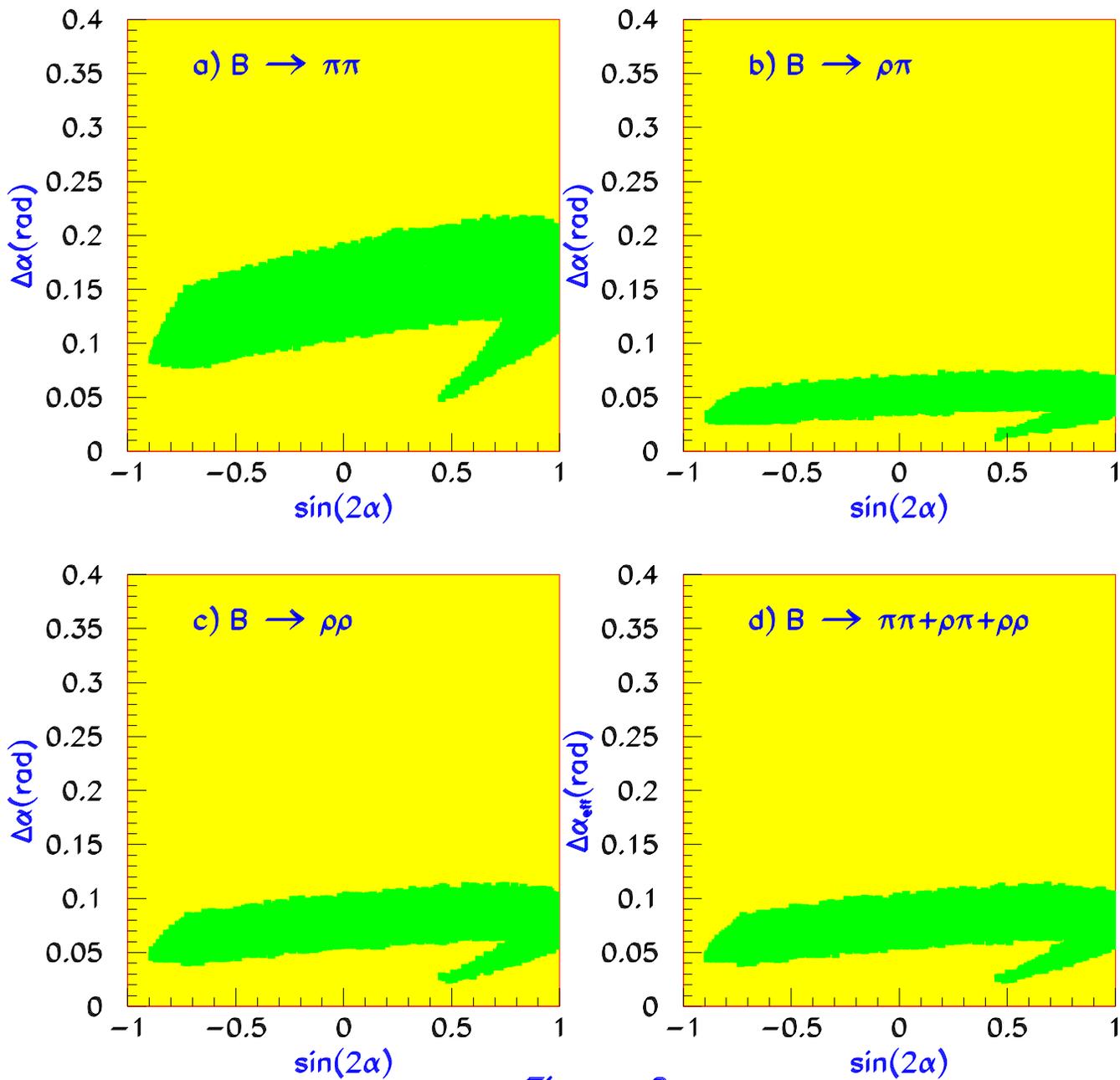
\item{\bf Fig. 2} The uncertainty $\Delta \alpha$ as a function of $\alpha$ for
the modes
$\pi^+\pi^-$, $\pi^+\rho^- + \pi^- \rho^+$ and $\rho^+\rho^-$ and for the sum
of all these modes.
\vskip 3 mm

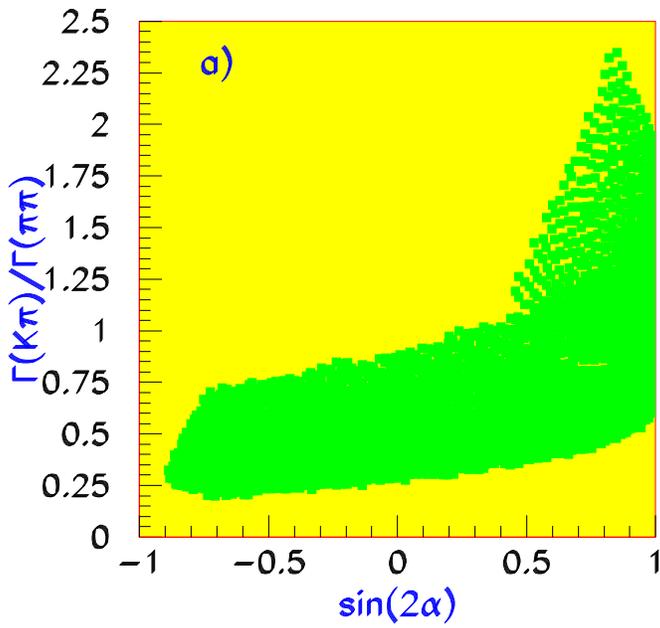
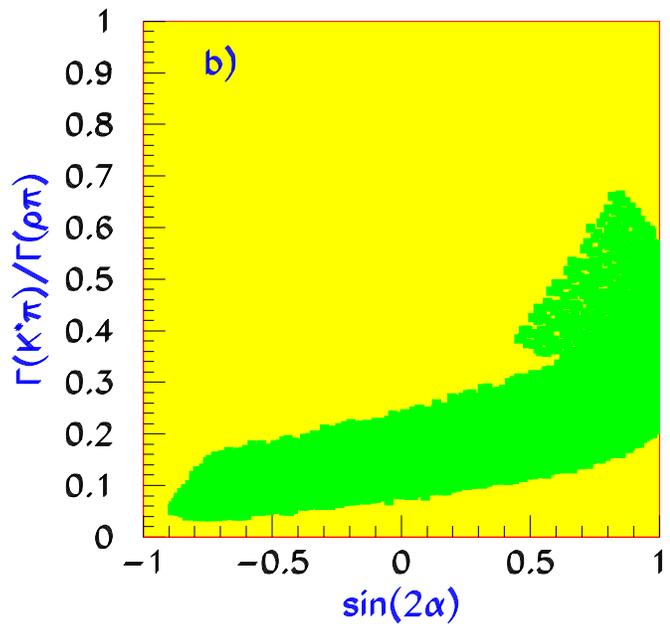
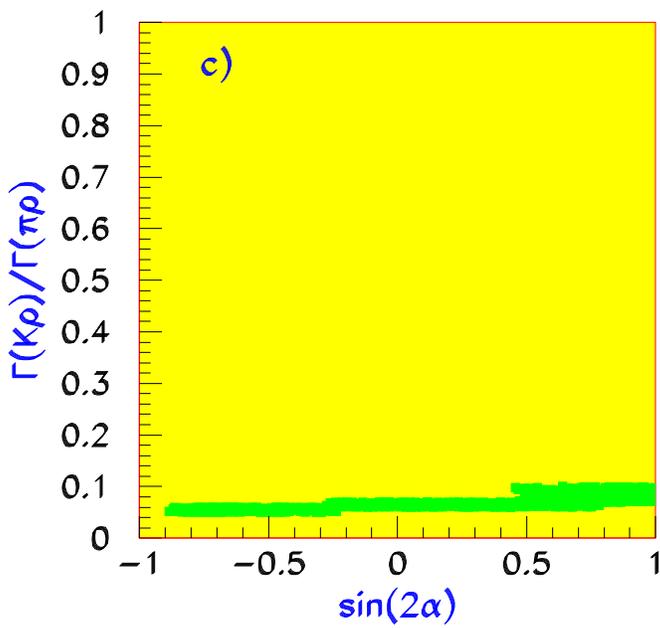
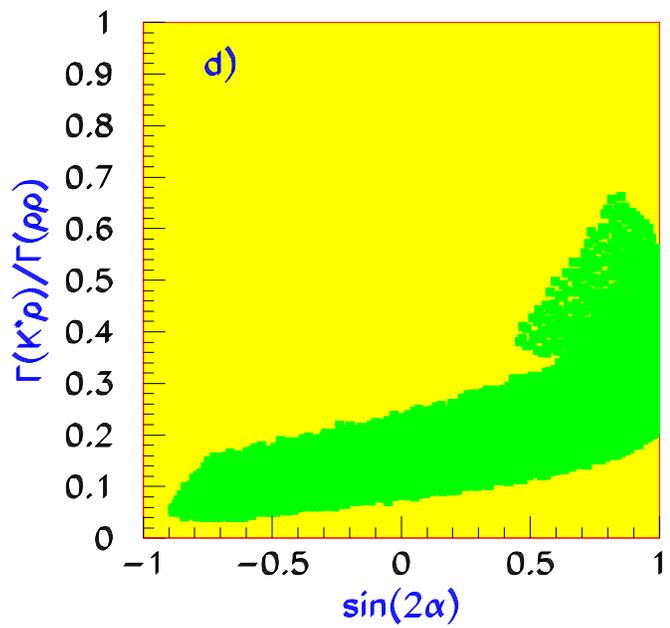
\item{\bf Fig. 3} The ratios of rates $\pi^+K^-/\pi^+\pi^-$,
$\pi^+K^{*-}/\pi^+\rho^-$,
$K^-\rho^+/\pi^-\rho^+$ and $\rho^+K^{*-}/\rho^+\rho^-$ as a function of $\sin
2 \alpha$. \vskip 3 mm

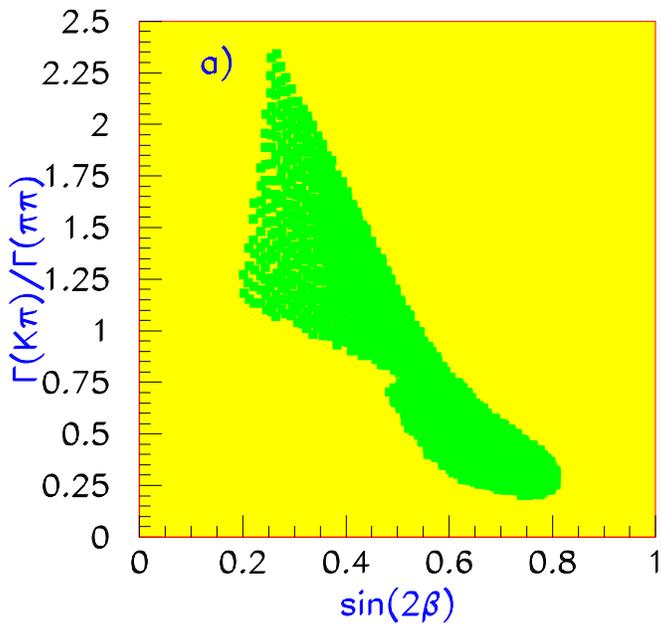
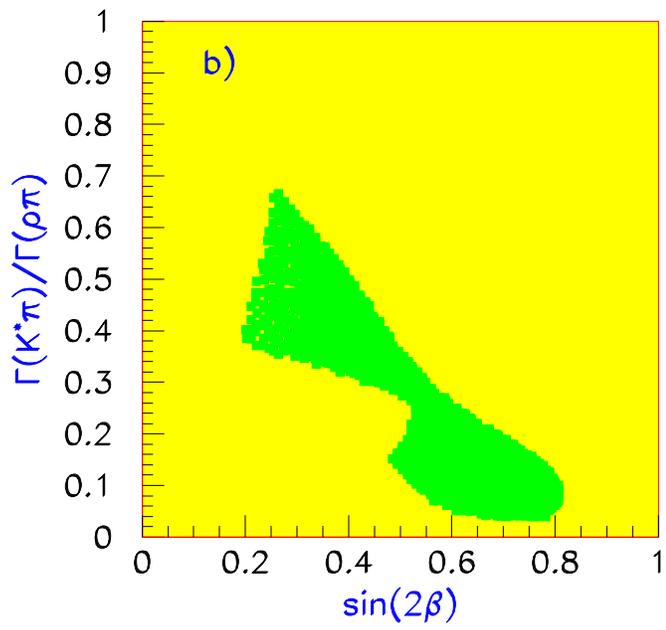
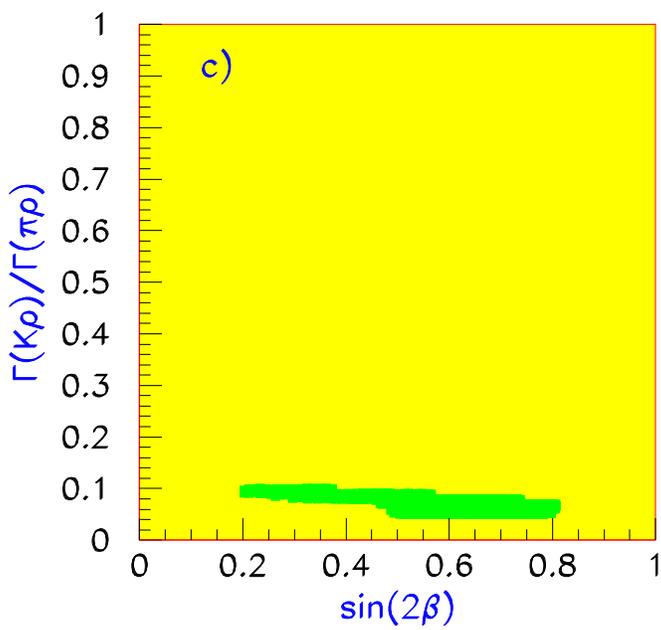
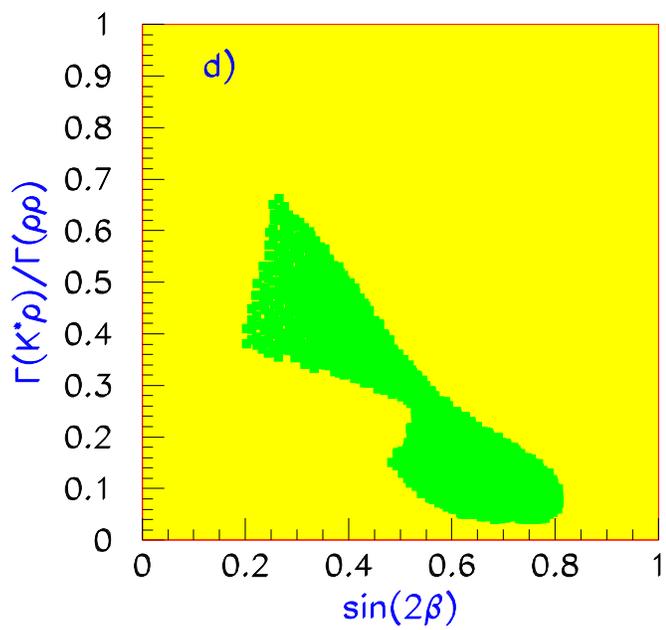
\item{\bf Fig. 4} The ratios of rates $\pi^+K^-/\pi^+\pi^-$,
$\pi^+K^{*-}/\pi^+\rho^-$,
$K^-\rho^+/\pi^-\rho^+$ and $\rho^+K^{*-}/\rho^+\rho^-$ as a function of $\sin
2 \beta$.
\par}

\vfill \supereject
\centerline{\bf Table 1}
\vskip 3 mm

$$\vbox{\offinterlineskip \halign{
\tv# &\omit\cc{#} &\tv# &\cc{#} &\tv# &\cc{#} &\tv# &\cc{#} &\tv#  \cr
\noalign{\hrule}
&\omit\cc{Decay mode} &&BR &&$D$ &&$\Delta \alpha$   & \cr
\noalign{\hrule}
&\omit\cc{$\bar{B}_d^0 \to \pi^+ \pi^-$}	&&1.94 $\times$ 10$^{-5}$	&&1	&&Arg $[
1 + 0.055
{1 - \rho + i \eta \over \rho - i \eta}]$ & \cr
\noalign{\hrule}
&\omit\cc{$\bar{B}_d^0 \to \pi^+ \rho^-$} &&1.53 $\times$ 10$^{-5}$ && && & \cr
\noalign{\hrule}
&\omit\cc{$\bar{B}_d^0 \to \rho^+ \pi^-$}	&&5.07 $\times$ 10$^{-5}$ && && & \cr
\noalign{\hrule}
&\omit\cc{$\bar{B}_d^0 \to \pi^ + \rho^- + \rho^+ \pi^-$}	&&6.60 $\times$
10$^{-5}$	&&0.84	&&Arg
$[ 1 + 0.019 {1 - \rho + i \eta \over \rho - i \eta}]$ & \cr
\noalign{\hrule}
&\omit\cc{$\bar{B}_d^0 \to \rho^+ \rho^- (L)$}	&&1.26 $\times$ 10$^{-4}$	&&1
&&Arg $[ 1 +
0.029 {1 - \rho + i \eta \over \rho - i \eta}]$ & \cr
\noalign{\hrule}
&\omit\cc{$\bar{B}_d^0 \to \rho^+ \rho^- (T, pv)$}	&&0.24 $\times$ 10$^{-5}$
&&1	&&Arg $[ 1
+ 0.029 {1 - \rho + i \eta \over \rho - i \eta}]$ & \cr
\noalign{\hrule}
&\omit\cc{$\bar{B}_d^0 \to \rho^+ \rho^- (T, pc)$} &&0.48 $\times$ 10$^{-5}$
&&-1	&&Arg $[
1 + 0.029 {1 - \rho + i\eta \over \rho - i\eta}]$ & \cr
 \noalign{\hrule}
&\omit\cc{$\bar{B}_d^0 \to \rho^+ \rho^- (T)$}	&&0.72 $\times$ 10$^{-5}$
&&-0.33 && & \cr
\noalign{\hrule}
&\omit\cc{$\bar{B}_d^0 \to \rho^+ \rho^-$} 	&&1.37 $\times$ 10$^{-4}$	&&0.93
&&Arg $[ 1 +
0.029 {1 - \rho + i \eta \over \rho -i \eta}]$ & \cr
\noalign{\hrule}
&\omit\cc{$\bar{B}_d^0 \to \pi^+ \pi^- + \pi^+ \rho^-$}	&&1.94 $\times$
10$^{-4}$
&&0.91 &&Arg $[ 1 + 0.029 {1 - \rho + i\eta \over \rho - i\eta}]$ & \cr
&\omit\cc{$+ \rho^+ \pi^- + \rho^+ \rho^-$} && && && & \cr
\noalign{\hrule}
}}$$

\vskip 3 mm
The uncertainty $\Delta \alpha$ and the dilution factor $D$ coming from Penguin
diagrams and the
branching ratios for the different modes ($L$ and $T$ denote longitudinal and
transverse~; $pc$ and
$pv$ mean parity conserving and parity violating).
 
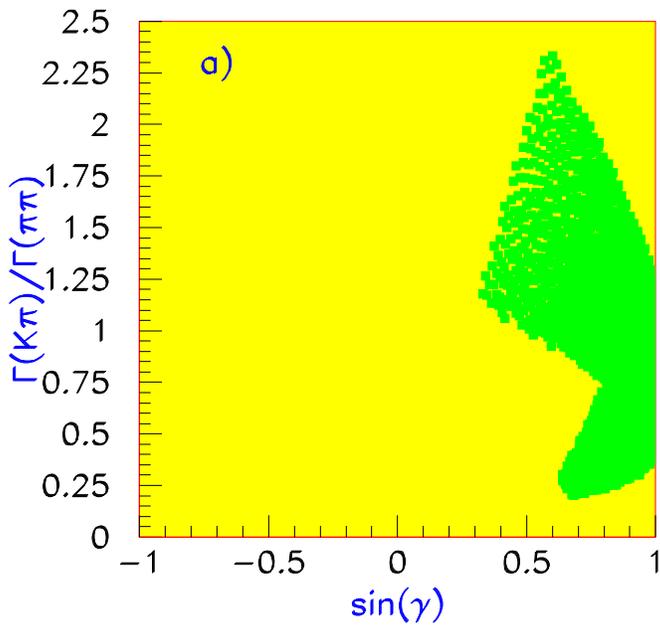
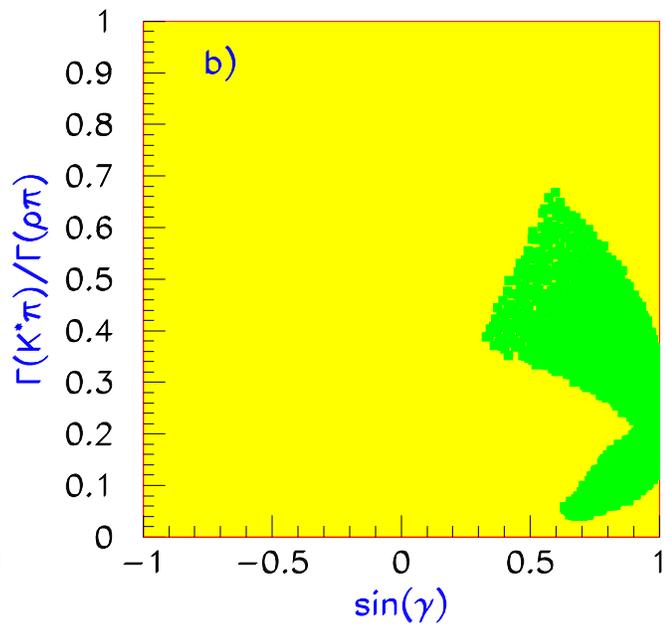
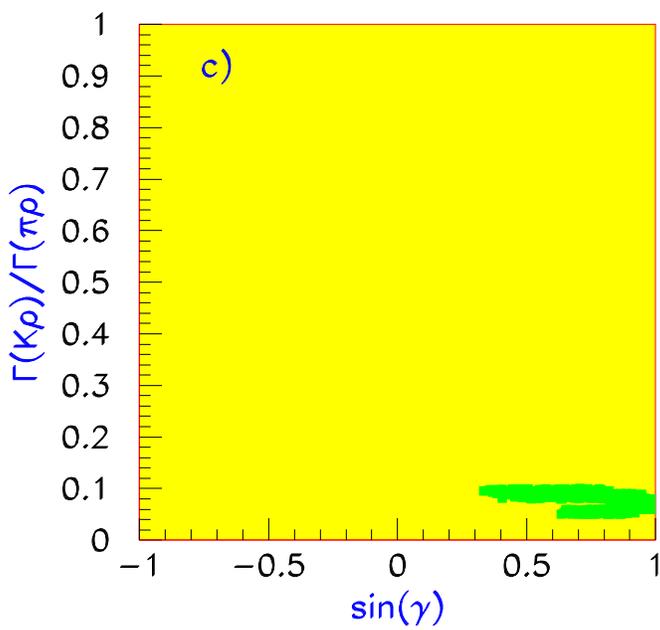
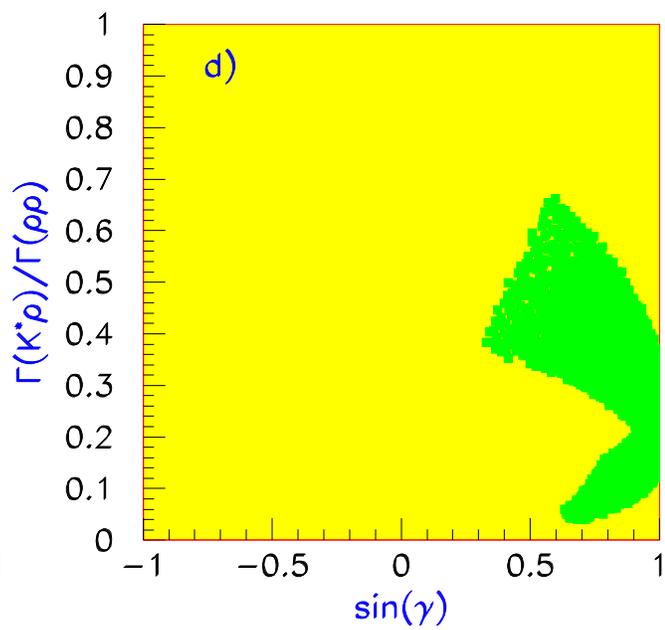
\bye